# Origin of the 60K plateau in $YBa_2Cu_3O_{6+x}$

V. M. Matic and N. Dj. Lazarov

*Laboratory of Theoretical Physics,
Institute of Nuclear Sciences "Vinca", 11001 Belgrade, Serbia*

**Abstract**

A model for charge transfer mechanism in $YBa_2Cu_3O_{6+x}$ high-$T_{ic}$ cuprate based on critical chain length concept is proposed to account for 60K and 90K plateaus in $T_c(x)$ dependence. It has been shown, when the statistics of CuO chain formation was described in terms of two dimensional asymmetric next-to-nearest neighbor Ising (ASYNNNI) model, that at any constant temperature below the top of OII phase there exists a uniquely defined value of critical chain length $l_{cr}(T)$ that yields a constant doping $p(x) \approx const$ over the regime of OII phase (related to 60K plateau of $T_c(x)$), while 90K plateau coincides with the monotonously increasing $p(x)$ over optimal doping level $p=0.16$ in the regime of OI phase. Short length chains ($l<l_{cr}(T)$), together with the first $l_{cr}(T)$-2 holes in longer chains ($l \geq l_{cr}(T)$), are taken as not capable of attracting electrons from $CuO_2$ planes. It is shown that only a part ($\approx 41\%$) of the remaining $l-l_{cr}(T)+1$ holes in the long chains can capture electrons. The results obtained indicate that the ASYNNNI model and two-plateaus-like behavior of $T_c(x)$ in $YBa_2Cu_3O_{6+x}$ are closely connected.

CuO$_2$ layers are key ingredients in all high-$T_c$ cuprates given the fact that superconductivity occurs in these materials when a part of 3$d$ copper electrons, usually between 5% and 27%, is taken away from the layers. The missing electrons are commonly referred to as "holes" which can move throughout the layers and make the material superconducting if the temperature is low enough. The electrons are conventionally subtracted from the layers either by chemical substitution of interlayer metal atoms, as for example, substitution of La$^{2+}$ by Sr$^{3+}$ in La$_{1-x}$(Sr)$_x$Cu$_2$O$_4$ and Y$^{2+}$ by Ca$^{3+}$ in Y$_{1-b}$(Ca)$_b$Ba$_2$Cu$_3$O$_{6+x}$, or by pumping oxygen into the material. Oxygen is introduced into separate layers in which it orders to form CuO chains and it is these chains that are known to act as efficient attractors of electrons from the CuO$_2$ layers. The number of created holes per Cu atom is typically denoted as "doping" $p$ and the notion that CuO$_2$ layers have an unavoidable role in the onset of superconductivity is further corroborated by the fact that a number of important physical characteristics, as, for example, the pseudogap energy $E_g$ and critical transition temperature $T_c$, are coupled to $p$ by universal relations that are common to practically all high-$T_c$ cuprates. Thus, it has been obtained empirically that the $T_c$ is scaled with $T_{c,max}$ (maximal transition temperature) through the following, approximately parabolic, function of the hole concentration [2]

$$T_c(p) = T_{c,\max}\left[1 - 82.6(p - 0.16)^2\right], \qquad (1)$$

which has its onset, maximum and termination at $p$=0.05, 0.16 (optimal doping) and 0.27, respectively.

YBa$_2$Cu$_3$O$_{6+x}$ superconductor has probably been the most thoroughly studied compound of all high-$T_c$ cuprates because it has a relatively simple synthesis route and it was the first superconductor discovered with $T_c$ above the liquid nitrogen temperature. The $T_c$ changes in nonlinear manner with oxygen composition revealing two well-known plateaus at 60K and 90K. These features are clearly associated with the Ortho-II and Ortho-I phases, respectively [3]. While the 90K plateau is in fact a broad maximum at $x\approx0.92$ that is associated with transition from underdoped to overdoped regime, the origin of 60K plateau at 0.5<$x$<0.7 is not yet well understood [3-5] (although, in fact, some advance has recently been made along this line [6]). One popular explanation is that when the oxygen content is increased over $x$=0.5, where Ortho-II phase emerges in the form of alternating columns of fully occupied and empty oxygen sites (directed along $b$ axes), additional oxygen fills the empty columns making a relatively small contribution to hole doping, inasmuch as they are far apart from each other and only a small fraction of them are able to form CuO chains that are long enough to initiate the charge transfer process. It has often been guessed that there might be a certain minimal (critical) chain length $l_{cr}$ defined so that only chains of length that is equal to, or greater than, $l_{cr}$ can effectively attract electrons from CuO$_2$ layers [3-6]. However, even though the existence of Ortho-II phase was resolved a long ago in terms of the classical two dimensional asymmetric next nearest neighbor Ising (ASYNNNI) model [7], no convincing explanation has been provided over the last 20 years as to why exactly the concentration $p(x)$ of holes induced in CuO$_2$ sheets would remain constant when $x$ changes within the Ortho-II phase regime, nor, if the critical chain length concept is presupposed, what the value of $l_{cr}$ would be equal to and how it can be determined.

Although it is well known that high-$T_c$ cuprates are complex quantum many-body systems with the pairing mechanism still remaining controversial, here we unambiguously grounds that it is the classical ASYNNNI model combined with the concept of minimal chain length (needed for charge transfer to take place) that accounts for constant doping at $p \approx 0.094$ in the region of 60K plateau, as well as for the broad maximum of $T_c$ at optimal doping ($p=0.16$) at $x \approx 0.92$.

There is a general agreement that copper in the chain (basal) plane can be either $Cu^{1+}$, which is the case when it is not coupled to the in-plane oxygen (but connected only to two apical O(4) ions and therefore 2-fold coordinated), or $Cu^{2+}$, when it is inserted within a CuO chain (4-fold coordinated), or at the chain end (3-fold coordinated). If isolated oxygen is introduced into the basal plane it then takes away two electrons from the two nearest neighbor $Cu^{1+}$ transferring them into $Cu^{2+}$ state. Thus, isolated oxygen does not have a tendency to attract an additional electron from other the parts of the system. When another oxygen is added to make a chain with two O atoms (chain of length $l=2$) there is only one electron available from its nearest neighbor copper coordination and the absence of another electron, needed for oxygen to become $O^{2-}$, is usually referred to as a "hole". In this way, a chain of length $l$ is seen as to have created $l-1$ chain-holes, which in principle can attract electrons from the other parts of the system, presumably from $CuO_2$ layers.

Since the state of quasi one dimensional electronic subsystem of a chain of a given length (say, $l$), which reflects the charge transfer effectiveness of the chain, is not expected to depend upon history of chain formation, one is free to assume that the chain has been formed by adding oxygen one by one as that would allow to shadow the evolution of charge transfer process as a function of $l$. Within the concept of critical chain length no charge transfer is supposed to occur unless $l=l_{cr}$ so that the first $l_{cr}-2$ initially created holes will stay inactive in attracting electrons from the planes. As $l$ further increases beyond $l_{cr}$ the transfer of charge is set on during which process the remaining $l-l_{cr}+1$ holes are created. Therefore, our strategy for counting doping is, first of all, to discard contribution not only of all holes in short chains ($l<l_{cr}$), but also of the first $l_{cr}-2$ holes in long chains ($l \geq l_{cr}$), for they had been created before any charge transfer took place (we shortly denote these holes as *passive* holes). It then naturally evolves that the number of attracted electrons (transferred holes) from the chain of length $l \geq l_{cr}$ should be as greater as more holes were created after the charge transfer process had been triggered (at $l=l_{cr}$), i.e. it should be proportional to $l-l_{cr}+1$ (we call these holes *the active* holes for their occurrence coincides with the development of charge transfer). In the case of Ortho-II phase, the concentration of active holes $h$ (the number of active holes per Cu) is given by

$$h = \frac{1}{4}\left[ n_{\alpha 1} \sum_{l=l_{cr}}^{\infty}(l-l_{cr}+1)f_{\alpha 1}(l) + n_{\alpha 2} \sum_{l=l_{cr}}^{\infty}(l-l_{cr}+1)f_{\alpha 2}(l) \right], \qquad (2)$$

where $n_{\alpha 1}$ and $n_{\alpha 2}$ denote the fractions of 3-fold coordinated Cu ions on two different sublattices of oxygen sites (usually denoted as $\alpha_1$ and $\alpha_2$), and $f_{\alpha 1}(l)$ and $f_{\alpha 2}(l)$ are corresponding fractions of CuO chains of the same length $l$.

If $N_{Cu}$ and $n=(n_{\alpha 1}+n_{\alpha 2})/2$ are total number and the fraction of 3-fold coordinated Cu in basal plane, then $(n/2)N_{Cu}$ is the total number of CuO chains and the number of passive holes per Cu is surely not greater than $(n/2)(l_{cr}-2)$, given the fact that no one of chains with $l<l_{cr}$ has more than $l_{cr}-2$ holes. At $x=1$ (OI stoichiometry) chains are very long, virtually infinite, and there are just a few chain ends in the system. This means that $n$ tends to zero as $x$ approaches to 1, so that the concentration of passive holes becomes negligible whatever the value of $l_{cr}$. This in turn implies that practically each chain ordered oxygen has created one active hole, i.e. $h≈1$ at $x≈1$. Since, in the $YBa_2Cu_3O_{6+x}$ system, one chain plane supplies holes to two $CuO_2$ layers, the doping would have been equal to $p=0.5$ if each active hole had succeeded to capture one electron. Experimental findings, however, clearly contradict such a scenario for it was found that $p(x≈1)$ only slightly overshoots 19% [2]. Thus, at least at $x≈1$, it can be stated $p=(\chi/2)h$ where $\chi$ ($≈40\%$), as defined by the number of effectively attracted electrons (transferred holes) per active hole, reflects the capability of an active hole to capture electron. On the other hand, long (infinite) chains also prevail at $x=0.5$ (OII stoichiometry) on every even column of oxygen sites. The concentrations of passive holes is also negligible here, bur $h$ cannot be greater that 0.5; in fact, $h$ can be only less than 0.5 due to thermally activated chain fragmentation (for example, one might expect that $h≈0.48$, as at $x≈0.5$, so in the region of the 60K plateau). Given the fact that 60K plateau of $T_c$, according to (1), corresponds to doping level $p≈0.094$ it appears that it is the same fraction (of nearly 40%) of active holes that is transferred to the layers not only at $x≈1$, but also at $x≈0.5$. The charge transfer model that we propose here assumes that the same percentage of active holes is transferred, not only at stoichiometries $x=0.5$ and $x=1$, at which long CuO chains are known to dominate, but as well at off-stoichiometry $0.5<x<1$ (and also at $x<0.5$) where chain fragmentation is more intense. We therefore propose that doping is connected to the active hole concentration $h$, as given by (2), by $p=(\chi/2)h$ ($\chi≈40\%$) throughout the whole range of oxygen concentration $0<x<1$. The quantity $\chi$ introduced in this way should be perceived as average capability of an active hole to capture an electron from the $CuO_2$ planes that lie above, or below, the basal plane (averaging is done over all chains in the system, or equivalently, over the whole volume of the material).

To calculate the hole concentration $h$ (eq. (2)) and doping $p=\chi h/2$, at a given point of $(x,T)$ space, it is necessary to determine the fractions of the 3-fold coordinated Cu and length distributions of CuO chains $f_{\alpha 1}(l)$ and $f_{\alpha 2}(l)$. We used Monte Carlo (MC) method applied to the ASYNNNI model to calculate these quantities for it is known that the model stabilizes both structures Ortho-II and Ortho-I that are responsible for 60K and 90K plateaus [9]. Although ASYNNNI model cannot stabilize other structures with longer periodicities along $a$ axis, like Ortho-III, Ortho-IV and Ortho-V, their superstructure reflections have already been reported to be much weaker than those of the main phases [8,9], so that they are thought to appear only as small patches embedded in large domains of main phases [10]. Besides, since these structures were mainly observed at oxygen compositions that correspond to transition region between the two plateaus it implies that the ASYNNNI model alone should account for both plateaus of $T_c(x)$, especially given the fact that none of these structures, except Ortho-II, was reported at $x<0.62$ [9]. The chain length distributions $f_{\alpha 1}(l)$ and $f_{\alpha 2}(l)$ were determined in the following way: In each MC step we counted the total numbers of chains $N_{\alpha 1}$ and $N_{\alpha 2}$, on sublattices $\alpha_1$ and $\alpha_2$, respectively ($N_{\alpha 1}$ and $N_{\alpha 2}$ are in fact equal to one half of unlike $V_2$ bonds on the corresponding $\alpha$ sublattices), as well as the numbers of chains of the same length, $N_{\alpha 1}(l)$ and $N_{\alpha 2}(l)$, for lengths ranging from $l=1$ to $l=300$. The ratios $N_{\alpha 1}(l)/N_{\alpha 1}$ and $N_{\alpha 2}(l)/N_{\alpha 2}$ were then equilibrated through the MC process and the so obtained values were finally assigned to $f_{\alpha 1}(l)$ and $f_{\alpha 2}(l)$. The MC calculations were performed using single-spin-flip Glauber dynamics, where the oxygen concentration $x$ is a functions of temperature $T$ and chemical potential $\mu$. We have studied lattices with periodic boundary conditions that consisted of 400x400 oxygen chain sites (O(1) sites, that split into two nonequivalent sublattices $\alpha_1$ and $\alpha_2$, in OII phase), and as many sites on β sublattice (O(5) sites). One MC step included flipping of all 2X(400X400) lattice spins and one MC run (at a particular point $(x, T)$) typically consisted of $3 \cdot 10^4$ to $5 \cdot 10^4$ MC steps, where only every tenth was used to calculate chain length distributions $f_{\alpha 1}(l)$ and $f_{\alpha 2}(l)$, $l=1,2, \ldots, 300$, and other relevant quantities (oxygen sublattice occupancies $x_1$ and $x_2$, 3-fold Cu fractions $n_1$ and $n_2$, etc.). At a certain number of points we have even used a really large number of MC steps, ranging from $10^5$ to $3 \cdot 10^5$.

At all calculated points of $(x,\tau)$ space ($\tau$ is a quantity that scales with $T$ according to $\tau=k_B T/V_1$, where $k_B$ is Boltzman constant and $V_1$ the nearest neighbor O-O interaction of the ASYNNNI model) it was obtained [11]

$$f_{\alpha i}(l) = \frac{1}{l_{av,\alpha i}} \left(1 - \frac{1}{l_{av,\alpha i}}\right)^{l-1}, \quad i=1,2 \quad , \tag{3}$$

where $l_{av,\alpha i}$ ($i=1,2$) denotes the average chain length on the corresponding sublattice ($\alpha_1$ or $\alpha_2$). Such a behavior of length distributions $f_{\alpha 1}(l)$ and $f_{\alpha 2}(l)$ was subsequently explained theoretically analyzing microscopic features of the ASYNNNI model lattice configurations [12]. In brief terms, the $l$ dependence of probability of a chain to have particular length $l$ can be derived in the following way: Consider a sequence of $N_y$ oxygen chain sites that are aligned along $b$ axis ($N_y$ is a large number, and the sites are connected by copper mediated $V_2<0$ bonds). Let $xN_y$ denotes the number of oxygen atoms on this column of $\alpha$ sites and let $nN_y$ stands for the number of unlike $V_2$ bonds ($n$ is therefore the fraction of 3-fold coordinated Cu along the column, and, consequently, $2x/n$ equals to the corresponding average chain length, $l_{av}$). These $xN_y$ oxygen atoms are generally divided into $(n/2)N_y$ groups (chains) that can have various lengths $l=1,2,\ldots$ . It is useful to recall that an each chain has two ends: one that is oriented towards positive side of $b$ axes (the "positive" end) and the other one, oriented towards negative $b$ axes (the "negative" end). Among these $xN_y$ oxygen atoms there are $(n/2)N_y$ of them that are located at the positive chain end, and thus the probability for an oxygen to be lying at the positive chain end is equal to $\omega=n/2x=\{l_{av}\}^{-1}$. Consequently, 1-$\omega$ is probability for oxygen to be located either within the interior of the chain, or at the negative chain end. Assuming that chains are created by adding oxygen one by one, starting from the negative end, one arrives at the conclusion that the probability for obtaining chain of length $l$ is equal to $f(l)=\omega(1-\omega)^{l-1}$ [11,12] {such a form of $f(l)$ dependence is known in the theory of probability as "geometric" probability distribution [13]). It should be noted, however, that a deeper analysis shows that the above reasoning applies only if fluctuations of energy of the ASYNNNI model are not too large [12]. Indeed, a certain deviation from linear behavior of $ln[f(l)]$ versus $l$ dependence has been found in the vicinity of the second order Ortho-I-to-Ortho-II phase transition curve (at $x>0.5$) [12] but, fortunately, such departures were observed only in a relatively narrow intervals $\Delta x \approx 0.07$ around critical points. Furthermore, our extended analysis (not shown here) shows that in the critical regime these deviations, of calculated $f_{\alpha 1}(l)$ and $f_{\alpha 2}(l)$ dependences from the expected behavior, were in a certain way compensated by summations in (2), so that the calculated $h(x)$ dependences were obtained to vary smoothly over the transition region at all $\tau$=const.

The so obtained values of length distributions $f_{a1}(l)$ and $f_{a2}(l)$ for $l=1,…,300$ were inserted into (2) to calculate concentration $h$ of active holes at different points $(x,T)$. The geometric-like behavior of $f(l)$ dependences ensures rather fast convergence of sums in (2). It should be mentioned, however, that the specific form of length distributions (3) makes it possible, instead of evaluating summations in (2) by first 300 terms, to transform each of sums into a closed analytical form, so that $h$ would be connected through analytical expression with average oxygen occupancies, $x_1$ and $x_2$, the 3-fold Cu fractions $n_1$ and $n_2$, and the parameter $l_{cr}$. Whatever the approach we used the calculated values of $h$ were obtained to be practically indistinguishable one from the other (even in the critical region of the ortho-I/ortho-II transition), but we nevertheless gave advantage to calculating the first 300 terms in estimating sums in (2), as we wanted to keep under control the departures of length distributions from (3) that are known to occur in the critical regime [12]. In addition, at each point of $(x,T)$ space $h$ was calculated for the whole range of values of cutoff parameter $l_{cr}$ spanning from $l_{cr}=1$ to $l_{cr}=50$, so that $h$ can be regarded as a function of three variables, i.e. $h=h(x,\tau,l_{cr})$ (aside from the fact that it also depends upon input parameters that define the ASYNNNI model, i.e. on O-O interactions $V_1>0$ (nearest neighbor) and $V_2<0$, $V_3>0$ (next nearest neighbor)). There is one remarkable feature of the hole concentration $h=h(x,\tau,l_{cr})$, as defined by (2), that we have found while thoroughly analyzing its behavior: whatever the magnitudes of interactions $V_1$, $V_2$, and $V_3$, when $h$ is considered as a function of $l_{cr}$ at different points $(x_i,\tau)$ that correspond to the same $\tau=const$ and oxygen concentrations $x_i$, $i=1,2,…$ spanning over the region of Ortho-II phase, all of these $h_{xi}(l_{cr})$, $i=1,2,…$ functions intersect at a single, well defined value of $l_{cr}$. This is shown in Figure 1 for interactions obtained by linear-muffin-tin orbital (LMTO) method [14] and at three different temperatures $\tau=0.45$, 0.38 and 0.30, but the similar behavior we have also obtained (not shown here) for the so-called "canonical" interactions $V_2=-0.5V_1$, $V_3=0.5V_1$. From Figure 1 it can be clearly seen that at a given $\tau=const$, the value of $l_{cr}$ at which all $h_{xi}(l_{cr})$ curves intersect depends on temperature in the way that it increases with the temperature decrease. Thus, at three temperatures $\tau=0.45$, 0.38, 0.30, the intersection values were found to be $l_{cr}=4(5)$, 6(7), 11(12), respectively. At a given temperature, the so obtained intersection value of $l_{cr}$ we name "the optimal minimal (critical) length" (denoted by $l_{cr,opt}(\tau)$) for it is the value at which $h(x)$ stays constant over the regime of Ortho-II phase. Such behavior of $h(x)$ at $\tau=const$, for the corresponding $l_{cr,opt}(\tau)$, is shown in Figure 2 at two temperatures: $\tau=0.45$ (Figures 2a and 2b, for $l_{cr,opt}(\tau)=4$ and 5, respectively) and $\tau=0.30$ (Figure 2c, for $l_{cr,opt}(\tau)=12$). From these results it can be seen that indeed $h(x)$ demonstrates a constant section at $x>0.5$ that is even more pronounced at lower temperatures.

Calculated $h(x)$ dependences were used to obtain doping versus $x$ dependences, $p(x)=\chi h(x)/2$, that were then inserted into (1) to yield corresponding $T_c(x)$s (Figures 3a-c). The parameter $\chi$ was varied slightly around its expected value $\approx 40\%$ [2] to achieve a better correlation between the so obtained $T_c(x)$'s and those from experiments [15] (shown by a solid line). We have indeed obtained $h$ in the plateau regime to be slightly lower than 0.5: $h=const=0.467$ (0.450) at $\tau=0.45$ for $l_{cr}=4$ ($l_{cr}=5$), $h=const=0.483$ at $\tau=0.38$ for $l_{cr}=7$, and $h=const=0.495$ at $\tau=0.30$ for $l_{cr}=12$. This gives $\chi=40.33\%$, 41.84%, 38.98%, 3804% for $h=0.467$, 0.450, 0.483, 0.495, respectively. Thus, not only experimental data on doping at $x\approx 1$ [2], but also the analysis at $x\approx 0.5$ (that falls into the regime of 60K plateau) both infer that $\chi$ is lying at some point between ≈38% and ≈42%. It should be noted that this result disagrees with estimations of Gawiec et al. [16,17] that long chains release up to 70% of their holes.

Although we used here the ASYNNNI model interactions $V_1$, $V_2$, and $V_3$ as obtained by Sterne and Wille [14] the issue of magnitudes of these interactions is still open so that new values were subsequently suggested [10]. Regardless of that, it seems very well established that nearest neighbor interaction $V_1$ should be ranking around 6.9$mRy$ [10,14], which fixes scaling between $T$ and $\tau$ to $\Delta\tau$≈0.1 $\Leftrightarrow$ $\Delta T$≈100K. On the other hand, as one of the most important features of $YBa_2Cu_3O_{6+x}$ phase diagram is the location of the top of Ortho-II phase along $\tau$ axis (at $x$≈0.5) and that it may well be affected by the magnitudes of $V_2$ and $V_3$ (which in fact are not known precisely), perhaps the best strategy to estimate the reduced temperature $\tau$ corresponding to room temperature is to determine the "distance" (in units of $\tau$) between room temperature and the top of Ortho-II phase. According to experimental data the top of Ortho-II phase corresponds to ≈125-140ºC [5,9], while theoretically obtained phase diagram for the LMTO interactions [18] points at $\tau$≈0.58, thus making $\tau$≈0.45 a fairly reliable estimation of room temperature.

The established correlation between room temperature and $\tau$=0.45 renders $l_{cr,opt}(\tau)$ (that plays the role of $l_{cr}$ in (2)) 4, or 5 (Figure 1$a$). A better estimation seems to be 5 for it yields a somewhat more pronounced 60K plateau than $l_{cr}$=4 (Figures 3$a$ and 3$b$), despite the fact that $l_{cr}$=4 would be closer to $l_{cr}$=3 that was proposed in some theoretical studies [16,17,19]. Besides, $l_{cr}$=5 appears to be well correlated with $\chi$≈2/5 and with the basic idea lying in the background of the minimal (critical) chain length concept suggesting that one isolated hole cannot efficiently attract an electron, but only a combined effect of several holes can achieve this goal. Both $l_{cr}$=5 and $\chi$≈2/5 imply that 3 chain-holes are still not enough to effectuate charge transfer, but that the joint impact of 5 holes suffices to attract two electrons (one per each $CuO_2$ layer).

In summary, despite the well known fact that in such highly correlated electron systems, as are the high-$T_c$ cuprates, the nature of the controversial pairing mechanism is genuinely quantum mechanical, we have shown here that certain aspects of their behavior can be explained in terms of classical models. Such is the classical ASYNNNI model that successfully accounts for unusual two-plateaus-like behavior of $T_c(x)$ in $YBa_2Cu_3O_{6+x}$. The obtained $T_c(x)$ dependence is in a remarkable correlation with experiment [15] for both $l_{cr}$=4 and $l_{cr}$=5, although we believe the later value is more realistic. It should be also pointed out that the presented results on $T_c(x)$ dependence are in a qualitative agreement with some previous results on the same topic [20,21]. Generally, it can be expected that the capability of an active hole to attract an unpaired 3$d$ electron, as expressed by the value of $\chi$, should depend upon the density $\rho_e$ of available electrons immediately above (below) the chain (aside from the fact that $\chi$ should also depend on a certain coupling between chains and planes). Our recently obtained results on $T_c(x)$ in $Y_{1-b}(Ca)_bBa_2Cu_3O_{6+x}$ ($b$=0.2) system [22] seem to be lying along this line, since obtained $\chi$≈33% can be understood in the light of the fact that introduction of ≈20% of Ca has additionally increased doping, and therefore reduced $\rho_e$, so that $\chi$ attained a lesser value than in the parent $YBa_2Cu_3O_{6+x}$ compound.

# Acknowledgements

This work has been funded by the Serbian Ministry of Science and Technology through the Project 141014.

# Figure Captions

**Figure 1.** Calculated $h(l_{cr})$ dependences at $\tau=const$ for several values of oxygen composition $x$ that span the range of OII phase: a) $\tau=0.45=const$; $0.52<x<0.61$ (shown at the top), b) $\tau=0.38=const$; $0.52<x<0.62$ (shown in the middle section), and c) $\tau=0.30=const$; $0.52<x<0.67$ (shown at the bottom).

**Figure 2.** Calculated values of $h$ as a function of $x$ at $\tau=const$: a) $\tau=0.45=const$, for $l_{cr}=4$ (shown at the top), b) $\tau=0.45=const$, for $l_{cr}=5$ (shown in the middle section), and c) $\tau=0.30=const$, for $l_{cr}=12$ (shown at the bottom).

**Figure 3,** Calculated $T_c(x)$ dependences, using $h(x)$ dependences from the Figure 2 and thereafter obtained $p(x)=\chi h(x)/2$ dependences, that were then inserted into (1): a) $\tau=0.45=const$, for $l_{cr}=4$ and $\chi=39.4\%$ (shown at the top), b) $\tau=0.45=const$, for $l_{cr}=5$ and $\chi=41\%$ (shown in the middle section), and c) $\tau=0.30=const$, for $l_{cr}=12$ and $\chi=37.6\%$ (shown at the bottom).

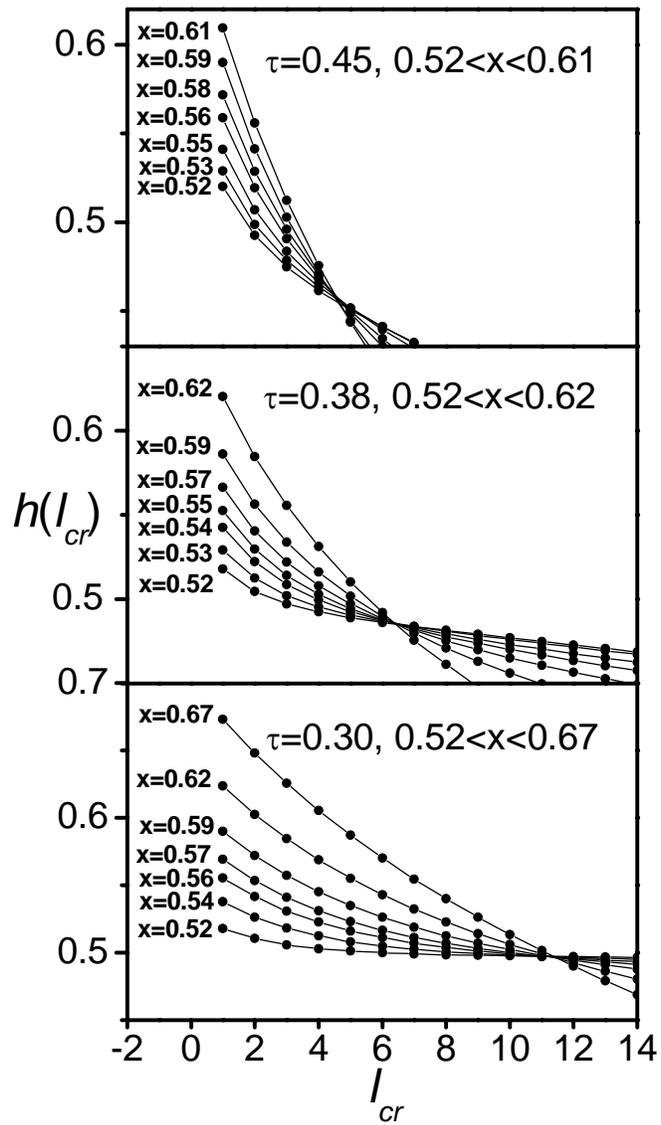

**Figure 1**

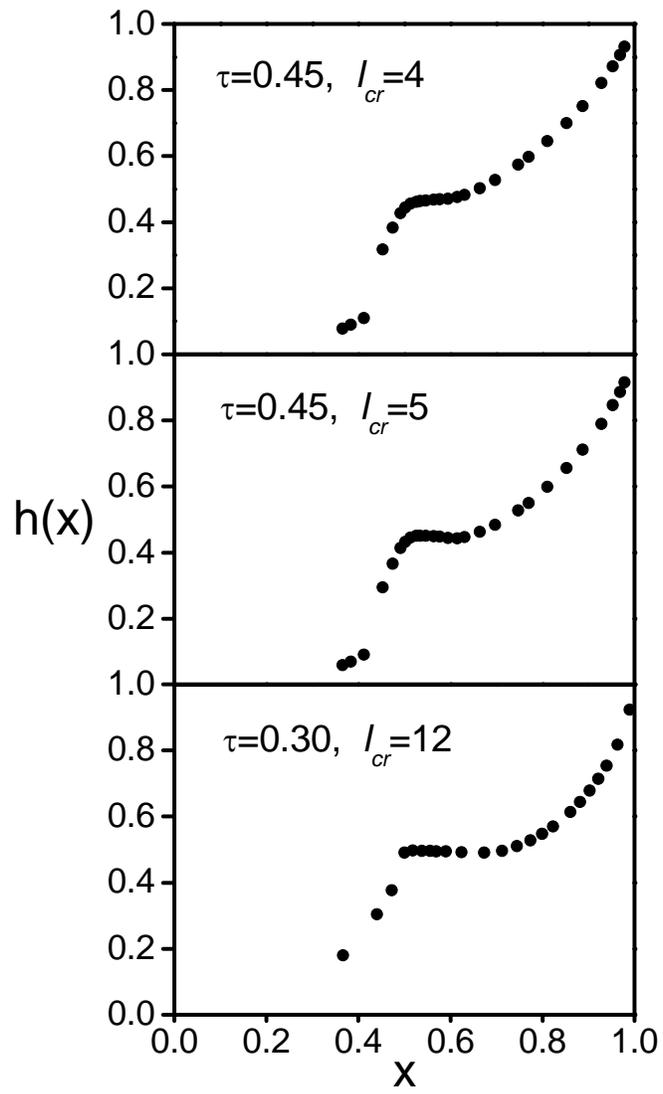

**Figure 2**

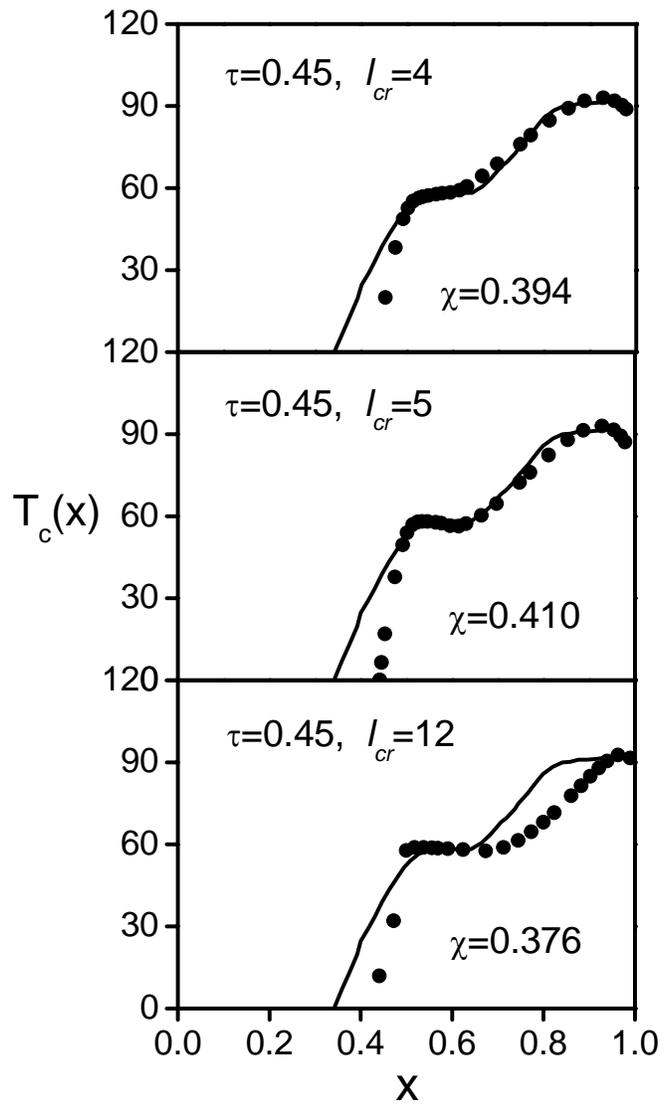

**Figure 3**